\documentclass[aps,twocolumn,superscriptaddress,prb]{revtex4}

\usepackage{graphicx}
\usepackage{revsymb}
\begin{document}


\title{GaMnAs-based hybrid multiferroic memory device}

\author{M. Overby}
\author{A. Chernyshov}
\author{L.~P. Rokhinson}
\affiliation{Department of Physics and Birck Nanotechnology Center, Purdue University, West Lafayette, Indiana
47907 USA}

\author{X. Liu}
\author{J.~K. Furdyna}
\affiliation{Department of Physics, University of Notre Dame, Notre Dame,
Indiana 46556 USA}

\date{\today}

\maketitle
\hyphenation{GaMnAs}

{\bf A rapidly developing field of spintronics is based on the premise that
substituting charge with spin as a carrier of information can lead to new
devices with lower power consumption, non-volatility and high operational speed
\cite{prinz98,wolf01}. Despite efficient magnetization
detection\cite{baibich88,binasch89,moodera95}, magnetization manipulation is
primarily performed by current-generated local magnetic fields and is very
inefficient. Here we report a novel non-volatile hybrid multiferroic memory
cell with electrostatic control of magnetization based on strain-coupled GaMnAs
ferromagnetic semiconductor\cite{ohno98} and a piezoelectric material. We use
the crystalline anisotropy of GaMnAs to store information in the orientation of
the magnetization along one of the two easy axes, which is monitored via
transverse anisotropic magnetoresistance\cite{tang03}. The magnetization
orientation is switched by applying voltage to the piezoelectric material and
tuning magnetic anisotropy of GaMnAs via the resulting stress field.}

In magnetic memories the information is stored in the orientation of
magnetization. The major weakness of current MRAM implementations lies in the
inherently non-local character of magnetic fields used to flip ferromagnetic
domains during the write operation. Neither thermal switching\cite{kerekes05}
nor current induced switching\cite{myers99} can solve the problem completely.

\begin{figure}[t]
\def\ffile{schematic}
\vspace{0.3in}
\includegraphics[scale=0.5]{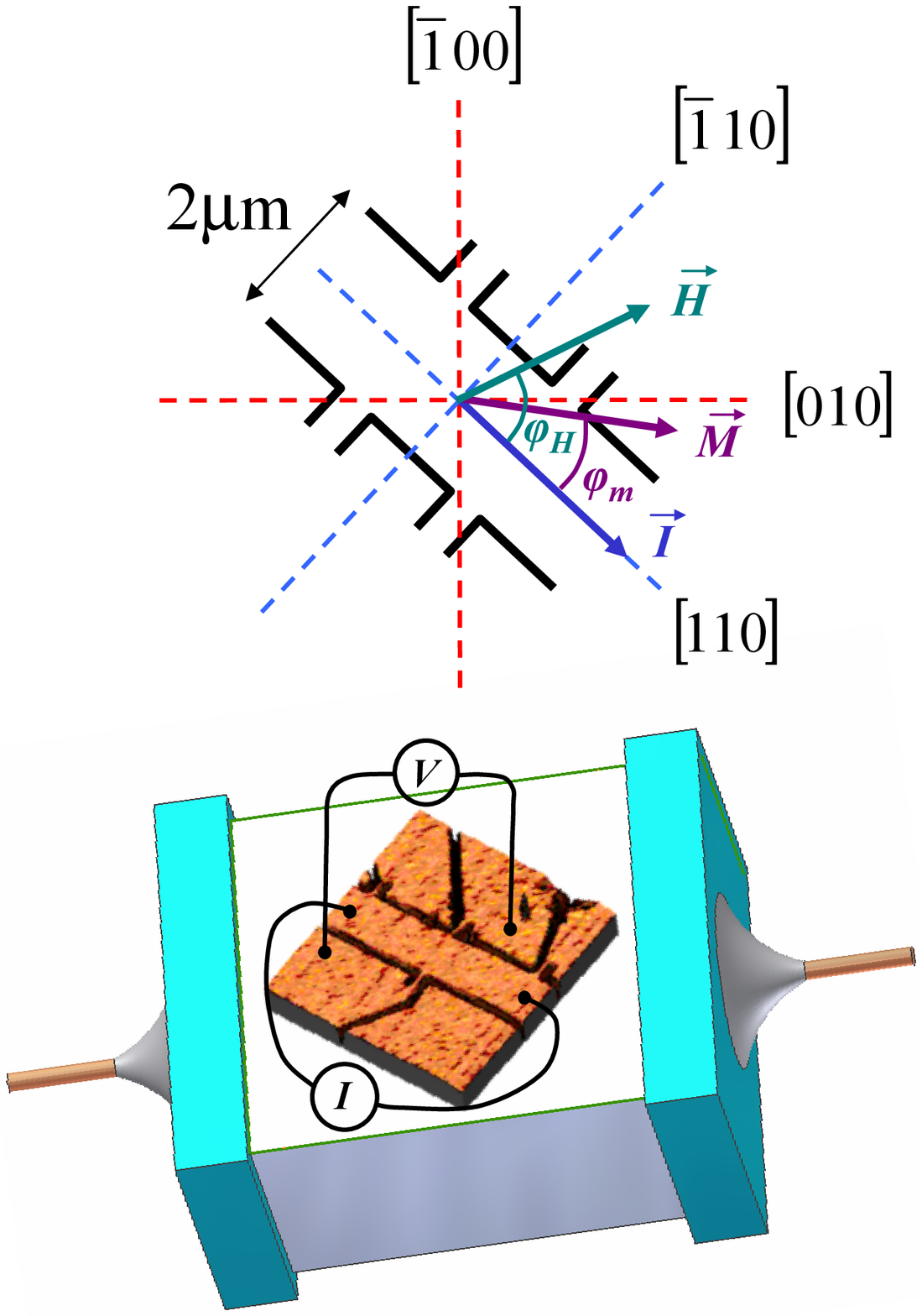}
\caption{ Top: sketch of the Hall bar with relative orientation of electrical current $\vec{I}$, magnetic
field $\vec{H}$ and magnetization $\vec{M}$. Bottom: an AFM image of a
$2\mu$m-wide Hall bar is schematically shown attached to the piezoelectric.
Strain is applied to the sample along [100] and [010] directions, red dashed
lines on the sketch.}
\label{\ffile}
\end{figure}

An attractive alternative to conventional ferromagnets are multiferroic
materials\cite{eerenstein06}, where ferromagnetic and ferroelectric properties
coexist and magnetization can be controlled electrostatically via
magnetoelectric coupling. In single-phase and mixed-phase multiferroics
ferromagnetic material must be insulating in order to avoid short-circuits.
Alternatively, magnetoelectric coupling can be introduced between ferromagnetic
and ferroelectric materials indirectly via strain, and in this case the
ferromagnetic material can be conducting. A conceptual memory device made of
ferromagnetic and piezoelectric materials has been proposed\cite{novosad00},
and permeability changes in magnetostrictive films\cite{arai94}, changes in
coercive field\cite{boukari07} and reorientation of the easy axis from in-plane
to out-of-plane in Pd/CoPd/Pd trilayers\cite{lee03} have been recently
demonstrated.

Some ferromagnetic materials have a complex anisotropic magnetocrystalline
energy surface and several easy axes of magnetization.  If the energy barrier
between adjacent easy-axis states can be controlled by strain, then a
multiferroic non-volatile multi-state memory device can be realized. Magnetic
anisotropy of dilute magnetic semiconductor GaMnAs films is largely controlled
by epitaxial strain\cite{welp03,liu03}, with compressive (tensile) strain
inducing in-plane (out-of-plane) orientation of magnetization. In GaMnAs
compressively strained to (001) GaAs there are two equivalent easy axes of
in-plane magnetization: along [100] and along [010] crystallographic
directions. Lithographically-induced unidirectional lateral relaxation can be
used to select the easy axis\cite{humpfner07,wenisch07}. In addition to the
large in-plane crystalline anisotropy there is an uniaxial anisotropy between
[110] and [1$\bar{1}$0] directions which is probably due to the underlying
anisotropy of the reconstructed GaAs surface\cite{welp04}.  We use this
magnetocrystalline anisotropy in combination with the magnetostrictive effect
to demonstrate a bi-stable memory device.

\begin{figure}[t]
\def\ffile{angleH}
\includegraphics[scale=0.9]{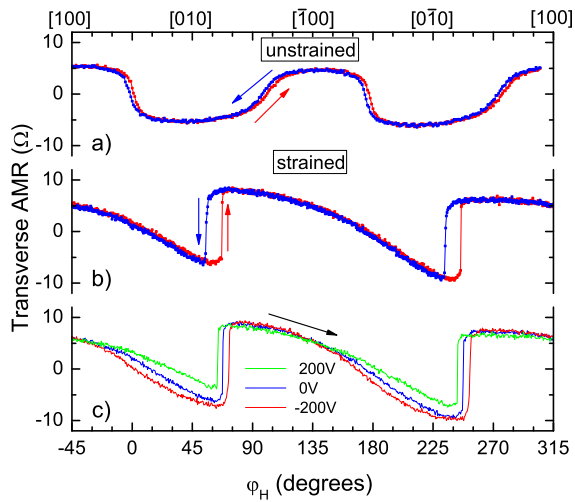}
\caption{Transverse AMR as a function of magnetic field angle $\varphi_H$
for unstrained (a) and strained (b,c) samples. Arrows indicate the field sweep
directions. Magnetic field is $H=50$ mT and $T=35$ K. Voltage applied to the
piezoelectric $V_{pzt}=0$ for the curves (b) and $V_{pzt}=\pm200$ V and 0 for
(c).}
\label{\ffile}
\end{figure}

Molecular beam epitaxy at 265$^\circ$ C was employed to grow 15-nm thick
epilayers of Ga$_{0.92}$Mn$_{0.08}$As on semi-insulating (001) GaAs substrates.
The wafers were subsequently annealed for 1 hour at 280$^\circ$ C in a nitrogen
atmosphere. Annealing increases the Curie temperature of the ferromagnetic film
to $T_c\sim 80$ K and reduces the cubic anisotropy. The GaMnAs layer was
patterned into 2 $\mu$m-wide Hall bars oriented along the [110] axis by
combination of e-beam lithography and wet etching, see Fig.~\ref{schematic}.
After lithography 3 mm x 3 mm samples were mechanically thinned to $\sim 100$
$\mu$m and attached to a multilayer PZT (piezoelectric lead-zirconium-titanate
ceramic)\cite{pzt} with epoxy, aligning the [010] axis with the axis of
polarization of the PZT. Application of positive (negative) voltage $V_{PZT}$
across the piezoelectric introduces tensile (compressive) strain in the sample
along the [010] direction, and strain with the opposite sign along the [100]
direction proportional to the piezoelectric strain coefficients $d_{33}\approx
-2d_{31}$. Both coefficients decrease by a factor of 15 between room
temperature and 4.2 K. The induced strain $\varepsilon=\Delta L/L$ for both
[010] and [100] directions was monitored with a biaxial strain gauge glued to
the bottom of the piezoelectric. Strain is proportional to the change of the
gauge resistance, and was measured in the Wheatstone bridge configuration:
$\Delta\varepsilon=\varepsilon_{[010]}-\varepsilon_{[100]}= (\Delta
L/L)_{[010]}-(\Delta L/L)_{[100]}=\alpha (R_{[010]}-R_{[100]})/R$, where
$\alpha$ is the gauge sensitivity coefficient and $R$ is the resistance of the
unstrained gauge. It has been shown before\cite{shayegan03} that the strain
gradient across the piezoelectric and the sample is negligible: i.e., gauges
glued on top of the sample and on the opposite side of the piezoelectric
measure similar strain.

\begin{figure}[t]
\def\ffile{switch}
\includegraphics[scale=0.85]{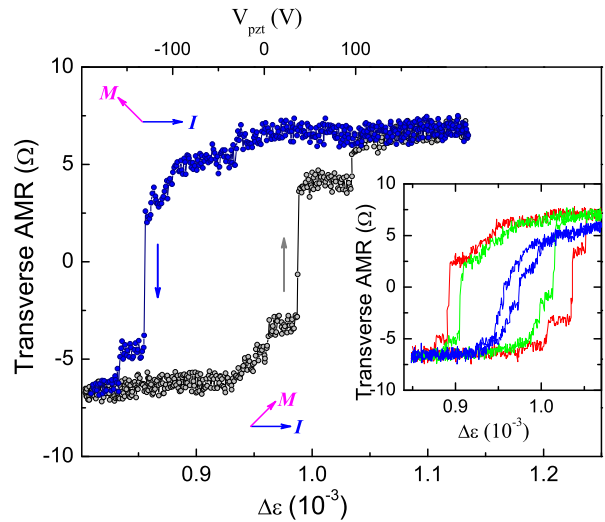}
\caption{Transverse AMR (TrAMR) for the strained Hall bar as a function of
uniaxial strain measured by a strain gauge. Static magnetic field $H=50$ mT is
applied at $\varphi_H=62^\circ$, the data is taken at $T=35$ K. On the top axis
approximate voltage on the piezoelectric is indicated. Orientation of
magnetization relative to the current flow is shown schematically for the
TrAMR$>0$ and TrAMR$<0$ states. In the inset TrAMR is plotted for $H=50$ mT
(red), 70 mT (green) and 100 mT (blue) measured at $T=25$ K.}
\label{\ffile}
\end{figure}

Direction of the in-plane magnetization $\vec{M}$ is measured via transverse
anisotropic magnetoresistance (TrAMR), also known as giant planar Hall
effect\cite{tang03}
\begin{equation}
TrAMR=(\rho_\|-\rho_\bot)\sin\varphi_m\cos\varphi_m,
\label{phe}
\end{equation}
where $\rho_\|$ and $\rho_\bot$ are the resistivities for magnetization
oriented parallel and perpendicular to the current. The sign and magnitude of
TrAMR depend on the angle $\varphi_m$ between magnetization $\vec{M}$ and
current $\vec{I}\|[110]$, see schematic in Fig.~\ref{schematic}. TrAMR reaches
minimum (maximum) when $\vec{M} \| [010]$ ($\vec{M} \| [\bar{1}00]$).
Longitudinal AMR is $\propto \cos^2\varphi_m$ and is not sensitive to the
magnetization switching between 45$^\circ$ and 135$^\circ$.

In Fig.~\ref{angleH} TrAMR is plotted for the strained and unstrained Hall bars
as magnetic field of constant magnitude $H=50$ mT is rotated in the plane of
the sample. For the unstrained sample (not attached to the piezoelectric) there
are four extrema for the field angle $\varphi_H$ around 45$^\circ$,
135$^\circ$, 225$^\circ$ and 315$^\circ$ with four switchings of magnetization
by 90$^\circ$ near [110] and [1$\bar{1}$0] directions, which reflects the
underlying crystalline anisotropy of GaMnAs. For the sample attached to the
piezoelectric there is a gradual change in the angle of magnetization with only
two abrupt $90^\circ$ switchings of the direction of magnetization per field
rotation, which indicates strong uniaxial anisotropy due to a highly
anisotropic thermal expansion coefficient of the PZT stack (+1 ppm/K along and
-3 ppm/K perpendicular to the piezoelectric stack). By applying a finite
voltage to the piezoelectric during cooldown we are able to control the
thermally-induced stress within $\Delta\varepsilon< 10^{-3}$ for $T<50$ K.

\begin{figure}[t]
\def\ffile{fit}
\includegraphics[scale=0.8]{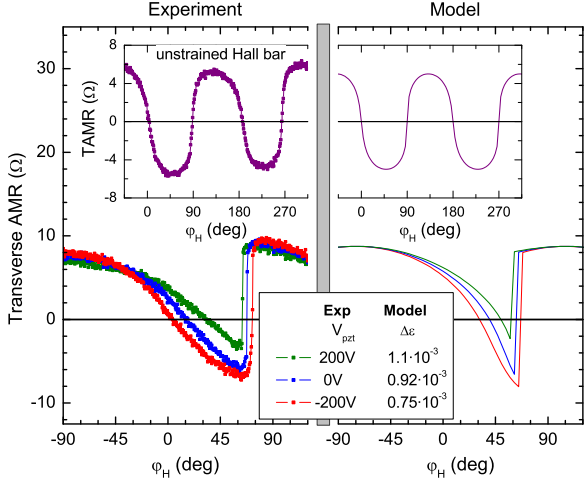}
\caption{ Experimentally measured (left panel) and modelled (right panel) transverse AMR
for the strained sample at different voltages on the piezoelectric (for
comparison data for an unstrained sample are shown in the inset). All data were
taken at $T=35$ K and $H=50$ mT. In the model we used experimentally measured
strain $\Delta\varepsilon$, other model parameters are discussed in the text.}
\label{\ffile}
\end{figure}

Application of $\pm200$ volts on the piezoelectric results in a shift of the
angle of magnetization switching by $\Delta\varphi_H\approx10^\circ$, which is
comparable to the size of the hysteresis loop in TrAMR vs $\varphi_H$ scans at
$V_{pzt}=0$, Fig.~\ref{angleH}(b,c). Applying a magnetic field of $H=50$ mT
oriented at $\varphi_H=62^\circ$, which corresponds to the middle of the
hysteresis loop, compensates for the thermally induced uniaxial strain
anisotropy and restores the original degeneracy between [010] and [$\bar{1}$00]
magnetization directions of the unstrained GaMnAs for $V_{pzt}\approx0$. An
additional strain is then applied by varying voltage on the piezoelectric. In
Fig.~\ref{switch} TrAMR is plotted as a function of measured strain $\Delta
\varepsilon$, the corresponding $V_{pzt}$ are approximately marked on the top
axis (there is a small hysteresis in $\Delta \varepsilon$ vs $V_{pzt}$). As
additional compressive strain is applied along [010], this direction becomes
the easy axis of magnetization and magnetization aligns itself with [010].  As
additional tensile strain is applied along [010] direction, the [$\bar{1}$00]
direction becomes the easy axis and polarization switches by 90$^\circ$. The
switching occurs in a few steps, indicating a few-domain composition of our
device. At $V_{pzt}=0$ the magnetization has two stable orientations,
$\vec{M}\|[\bar{1}00]$ and $\vec{M}\|[010]$, and the orientation can be
switched by applying a negative or a positive voltage on the piezoelectric.
Thus, the device performs as a bi-stable non-volatile magnetic memory with
electrostatic control of the state.

The center of the loop can be shifted by adjusting $\varphi_H$: e.g., a
1$^\circ$ change shifts the center of the loop by $\Delta\varepsilon\approx
3.5\cdot 10^{-5}$. As $H$ increases, the size of the hysteresis loop decreases,
and the hysteresis vanishes for $H>100$ mT; see inset in Fig.~\ref{switch}. At
$H<40$ mT the loop increases beyond the $\pm200$ piezoelectric voltage span. In
our experiments the magnetic field balances the residual strain due to
anisotropic thermal expansion of the PZT. Alternatively, intrinsic
piezoelectric properties of GaAs can be utilized, in this case there will be no
thermally induced strain and electrostatic switching of the magnetization
direction can be realized without applying an external compensating magnetic
field. Scaling of the piezoelectric element from 0.5 mm down to $\sim1\ \mu$m
will compensate for a small strain coefficient in GaAs ($d_{33}^{pzt}\approx
10\cdot d_{14}^{GaAs}$ at 30 K), reduce operating voltage to a few volts and
allow electrostatic control of individual memory cells.

\begin{figure}[t]
\def\ffile{sim}
\includegraphics[scale=0.95]{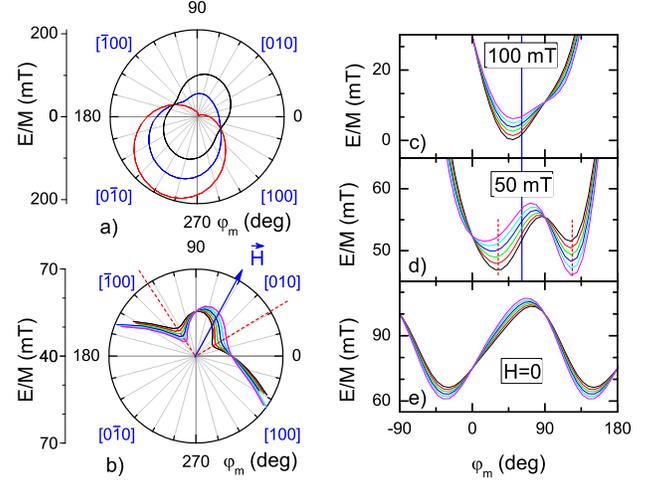}
\caption{ a) Polar plot of magnetostatic energy $E/M$ as a function of magnetization
angle $\varphi_m$ for $H=0$ (black), 50 mT (blue) and 100 mT (red) for
$\varphi_H=62^\circ$ and $V_{pzt}=0$ ($\Delta\varepsilon K_\varepsilon=16$ mT).
b-d) Angular dependence of $E/M$ for $\Delta\varepsilon K_\varepsilon=13-19$ mT
(black to magenta) for $H=0$ (e), $H=50$ mT (b,d) and $H=100$ mT (c). Blue line
and arrow mark $\varphi_H=62^\circ$, dashed red lines indicate two stable
orientations of magnetization.}
\label{\ffile}
\end{figure}

We model the strain along [100] and [010] directions as an extra magnetostatic
energy density term $2\varepsilon_{[100]}
K_{\varepsilon}\sin^2(\varphi_m+45^{\circ})+2\varepsilon_{[010]}
K_{\varepsilon}\sin^2(\varphi_m-45^{\circ})=\Delta\varepsilon
K_{\varepsilon}\sin(2\varphi_m)+const$. Then, for a single domain magnet the
free energy density can be written as
\begin{eqnarray}
E=&&K_u \sin^2(\varphi_m) + K_1/4\cos^2(2\varphi_m)\nonumber\\
&& + HM\cos(\varphi_m-\varphi_H)+\Delta\varepsilon
K_{\varepsilon}\sin(2\varphi_m)
\label{energy}
\end{eqnarray}
omitting constant offset, where $K_1$, $K_u$ and $K_{\varepsilon}$ are cubic,
uniaxial and strain anisotropy constants; $H$ is the applied in-plane magnetic
field; and $\varphi_m$ and $\varphi_H$ are the angles between [110] direction
and magnetization and magnetic field respectively; see schematic in
Fig.~\ref{schematic}. We assume that $K_{\varepsilon}$ is the same for [100]
and [010] directions.

In equilibrium the magnetization orientation $\varphi_m$ minimizes the free
energy, $dE/d\varphi_m=0$ and $d^2E/d\varphi_m^2>0$. The TrAMR can be
calculated from Eq.~\ref{phe} for a given angle $\varphi_H$ of the external
field $H$. From the fits to the experimental TrAMR data we can extract the
anisotropy constants $K_1$, $K_u$ and $K_\varepsilon$. The model captures all
the essential features of the data, and corresponding fits are shown in
Fig.~\ref{fit} for the strained and unstrained devices. For unstrained device
anisotropy fields $2K_1/M=40$ mT and $2K_u/M=6$ mT. These values are
significantly smaller than the previously reported values for as-grown (not
annealed) wafers\cite{tang03,shin07}. For the sample attached to the
piezoelectric the crystalline anisotropy field remains the same, but the
uniaxial anisotropy increases to $2K_u/M=50$ mT.  The strain-induced anisotropy
field $\Delta\varepsilon K_\varepsilon/M$ varies between 13 mT and 19 mT for
different $V_{pzt}$ between -200 V and 200 V, the coefficient
$K_\varepsilon/M=17$ T.

To illustrate the mechanism of magnetization switching we plot magnetic energy
density normalized by magnetization $E/M$ (Eq.~\ref{energy}) as a function of
$\varphi_m$ in Fig.~\ref{sim}. At $H=0$ there are only two minima along the
[100] axis due to the large uniaxial strain (see Fig.~\ref{sim} (a,e)), caused
by anisotropic thermal expansion coefficient of the piezoelectric.

With external magnetic field $H=50$ mT applied at $\varphi_H=62^\circ$ E/M has two
minima: at $\varphi_m=32^\circ$ and at $123^\circ$, i.e., close to [010] and to
[$\bar{1}$00] crystallographic directions. For the strain field $\Delta\varepsilon
K_\varepsilon/M=13$ mT the global minimum is at $\varphi_m=32^\circ$, and in equilibrium
the magnetization is oriented along the [010] direction. As the strain field increases to
19 mT the two minima switch, and $\varphi_m=123^\circ$ becomes the global minimum. It is
interesting to note that there is always a small barrier between the two minima. Unless
the barrier is an artifact of our model, the switching of magnetization should be either
temperature activated or should occur via macroscopic quantum tunnelling.

\bibliographystyle{revtex-nature}
\bibliography{rohi,dms-pzt}

{\bf Acknowledgements} The work was supported by NSF under the grants
ECS-0348289 (Purdue) and DMR-0603752 (Notre Dame).\\

\end{document}